\documentclass[graybox]{svmult}

\usepackage{type1cm}        
\usepackage{makeidx}         
\usepackage{graphicx}        
\usepackage{multicol}        
\usepackage[bottom]{footmisc}

\usepackage{newtxtext}       %

\usepackage{graphicx}
\usepackage{amsmath}
\usepackage{graphics}
\usepackage{epsfig}
\usepackage{amssymb}
\usepackage{hyperref}

\newcommand{\be}{\begin{eqnarray}}
\newcommand{\ee}{\end{eqnarray}}
\newcommand{\bea}{\begin{eqnarray*}}
\newcommand{\eea}{\end{eqnarray*}}

\def\X{{\Bbb X}}

\begin{document}


\title{Statistical Modelling  for Improving Efficiency of Online Advertising}
\titlerunning{Statistical Modelling  for Online Advertising}

\author{Irina Scherbakova and Andrey Pepelyshev %
 and Yuri Staroselskiy and Anatoly Zhigljavsky and Roman Guchenko}
\authorrunning{I Scherbakova, A Pepelyshev, Y Staroselskiy, A Zhigljavsky, R Guchenko}
%

\institute{Irina Scherbakova, Yuri Staroselskiy, Roman Guchenko\at Crimtan, London, UK \email{ishcherbakova@crimtan.com},\email{yuri@crimtan.com},\email{rguchenko@crimtan.com}
\and
Andrey Pepelyshev, Anatoly Zhigljavsky \at Cardiff University, Cardiff, UK \email{pepelyshevan@cardiff.ac.uk},\email{ZhigljavskyAA@cardiff.ac.uk} }

\maketitle

\abstract{
Real-time bidding has transformed the digital advertising landscape, allowing companies to buy website advertising space in a matter of milliseconds in the time it takes a webpage to load. Joint research between Cardiff University and Crimtan has employed statistical modelling in conjunction with  machine-learning techniques on big data to develop computer algorithms that can select the most appropriate person to which an ad should be shown. These algorithms have been used to identify  suitable bidding strategies for that particular advert in order to make the whole process as profitable as possible for businesses.
Crimtan's use of the algorithms have enabled them to improve the service that they offer to clients, save money, make significant efficiency gains and attract new business. This has had a knock-on effect with the clients themselves, who have reported an increase in conversion rates as a result of more targeted, accurate and informed advertising.
We have also used mixed Poisson processes for modelling for analysing repeat-buying behaviour of online customers.
To make numerical comparisons, we use real data collected by Crimtan in the process of running  several recent ad campaigns.
}


\subsection*{Introduction}

The work is devoted to  developing models of consumer behaviour for improving efficiency of  internet advertisement, where
 fast decisions have to be made on whether to show a given ad to a particular user.
Such decisions are based on
information extracted from big data sets containing records of previous impressions, clicks and subsequent purchases.

During the last two decades online advertisement grew  enormously, evolved into a major multi-billion dollar industry and   became the most significant part of the total advertisement market.
The market has been driven by the growth of programmatic buying (when decisions are made by computers  rather than people), and more specifically the process of real-time bidding, in which companies are given the opportunity to buy advertising space on websites as part of bidding process in a 'virtual auction'. This process occurs in the milliseconds that it takes for a webpage to load and is repeated billions times each day.
Demand partners typically collect databases with logs of all previous requests from auctions,
impressions, clicks, conversions and users who visited a website which is currently advertised.
These logs typically contain an anonimized user id, a browser name, an OS name, a geographical information
derived from the IP address and a webpage link where an auction is run.
Merging these datasets with third party data sources provides possibilities for contextual, geographical and behavioural targeting.

We consider the problem of online advertisement via auctions holding by
independent ad exchanges from the position of a demand partner which wants to
optimise the conversion rate. The demand partner has to decide how reasonable is it  showing an ad in regard to a request from an auction
and then possibly suggest a bid. The process of showing online advertisements through the so-called Real-Time Bidding (RTB) systems occurs  billions of times every day (see \cite{Paper5} for a detailed description of an RTB system).

The demand partner has to solve the problem
of maximizing either
the click through rate (CTR) or the conversion rate
by targeting a set of requests
under several constraints:
(a) budget (total amount of money available for advertising),
  (b) number of impressions $N_{total}$ (the total amount of ad exposures), and
  (c) time (any ad campaign is restricted to a certain time period).


In the rest of the paper, we discuss several issues related to (i) development of specialized machine learning (ML) algorithms for  online advertising with addressing the problem of assessing  the significance of different factors on probabilities of visits to particular sites, clicks and conversions
and (ii) adaptation of the mixed Poisson process model for modelling consumer/user behaviour  in internet environment.

\subsection*{Development of specialized ML algorithms for  online advertising and comparative studies}

In  \cite{PSZ2015,PSZ2016,Paper5} we have made a critical analysis of several procedures for on-line advertisement,
provided a unified point of view on these procedures and have had a close look at the so-called `look-alike' strategies.
In \cite{Paper5}, we have also  studied relative influence of different factors on the conversion rate and hence develop simple procedures which are
very computationally light but achieve the same accuracy as computationally demanding algorithms
like Gradient Boosting Machines (GBM) or Field-Aware Factorization Machines (FFM).
Note that the number of parameters in the simplest  FFM models is the  sum of all factor levels plus perhaps  interactions between factor levels. It counts to millions and if at least some interactions are taken into account then the count takes to much larger numbers.
Unlike standard ML models, the models developed in \cite{Paper5} have a relatively small number of parameters: for predicting the conversion rate, we have propose
 a sparse model where only a few most significant factors are used .

Formal statement of the problem is as follows.
Suppose that the advertisement  we want to show is given and first assume that the price for showing a given ad is fixed; we shall also ignore the time constraint. Then the
problem can be thought of as an optimization problem for a single optimality criterion which we choose as CTR.
We discuss a generic  adaptive targeting strategy which
should yield the decision whether or not to show  the ad to a request from a webpage visited by a user.
If the strategy decides to show the ad, it then has to propose a bid.
An adaptive decision should depend on the current dataset of impressions and clicks
which include all the users to whom we have shown the ad before and those who have  clicked on the ad.
Note that the dataset size  $N$ grows with time.
We can increase the size of the dataset by including all our previous impressions of the same or similar advertisements
(perhaps applying some calibration to decrease the influence of past ad campaigns),
so that $N$  could be very large.

Denote the $i$-th request by $X_i=(x_{i,1}, \ldots,x_{i,m})$, $i=1, \ldots, N$,
where $m$ is the number of features (factors); these features include the behavioural characteristics of the user, characteristics of the website, time of exposure, the device used (e.g. mobile telephone, tablet, PC , etc.).
We also assume that for any two points $X$ and $X^\prime \in \X$,
we can define a similarity measure $d(X,X^\prime)$
which does not have to satisfy mathematical axioms of the distance function.

If $\X$ is a discrete set with all possible requests $X=(x_{1}, \ldots,x_{m}) \in \X$  given on the nominal scale then we can use the Hamming distance
or
the weighted Hamming distance $d(X,X^\prime)=\sum_{j=1}^m w_j \delta(x_j,x'_j)$,
where the coefficients $w_j$ are positive and proportional to
the importance of the $j$-th feature (factor), $j=1,\ldots,m$.
These weight coefficients can be computed adaptively on the basis of the analysis
of previous data of similar advertising campaigns.

Alternative ways of defining the similarity measure $d(X,X^\prime)$
are a logistic model for $p_X$
(as is done in FFM),
or to use sequential splitting of the set $\X$ based on the values of
the most important factors of $X$ (as in GBM).
For FFM, the distance is defined on the space of parameters of the logistic model
but in GBM  $d(X,X^\prime)$ is small if $d(X,X^\prime)$ belongs
to the same subset of $\X$ and it is large if the subsets which $X$ and $X^\prime$
belong to have been split at early stages of the sequential splitting algorithm (that is, the values of the most influential features are very different).

To identify how the $i$-th factor affects  the conversion rate $p(X)$,
we use the concept of  mutual information.
To find the relative influence of the $i$-th factor in the sense
the mutual information based on the Shannon entropy,
we consider the statistic defined by
$
 I^{(Sh)}_i=\sum_{k=1}^{L_i} \sum_{s=0}^1
 p_{i,k,s}\log_2 {p_{i,k,s}}/{p_{i,k,\ast}\,p_{i,\ast,s}} \, .
$
To find the relative influence of the $i$-th factor in the sense
the mutual information based on the Renyi entropy of order $\alpha$,
we consider
$
 I^{(Re,\alpha)}_i= \log_2\sum_{k=1}^{L_i} \sum_{s=0}^1
 {p^{\alpha}_{i,k,s}}/{p^{\alpha-1}_{i,k,\ast}\,p^{\alpha-1}_{i,\ast,s}} \, .
$
For estimating the conversion rate $p(X)$,  we propose
$
 \hat p(X)={\sum_{i=1}^m I_i q_{i,k_i}}/{\sum_{i=1}^m I_i}
 \label{eq:pred-p}
$
where $k_i$ is such that $X_i=l_{i,k_i}$ and
$I_i$ is a relative influence of the $i$-th factor.
If we want to use a sparse predictive model then we can set the values of $I_i $ such that $I_i \leq \epsilon$ to zero, for some small $\epsilon>0$.

In \cite{Paper5} we provide careful comparison of the ML procedure that is based on the use mutual information for selecting  the  most important features
with the classical ones including GBM and FFM. We have shown that the main advantage of the proposed model  is its simplicity and time efficiency while the accuracy
of all considered ML procedures are practically identical.

\subsection*{Mixed Poisson processes for modelling dynamics of visits to websites, clicks and
conversions}

\begin{figure}[ht!]
\begin{center}
{\includegraphics[width=0.95\textwidth]{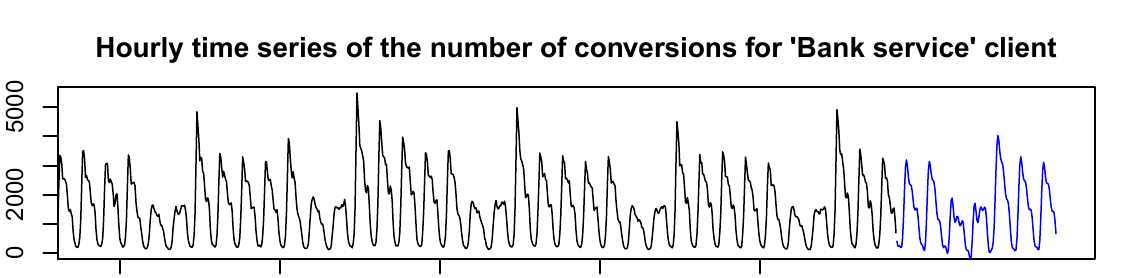}}\\
{\includegraphics[width=0.95\textwidth]{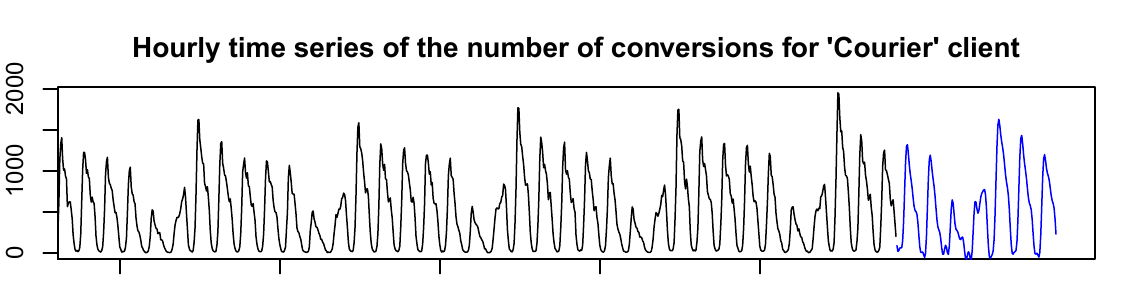}}\\

\end{center}
\caption{Typical hourly series for the the number of conversions (or simply visits to websites) and their forecasts (blue).  }
\label{fig:hourly}
\end{figure}

\begin{figure}[ht!]
\begin{center}
{\includegraphics[width=0.79\textwidth]{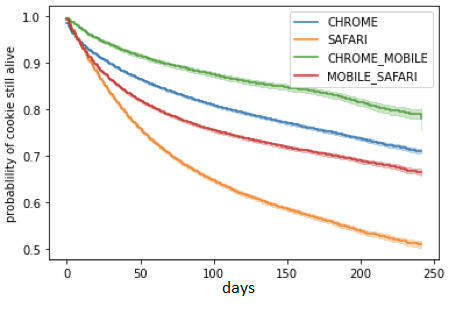}}
\end{center}
\caption{Cookies survival rates for different types of browsers.  }
\label{fig:cookies}
\end{figure}

\begin{figure}[ht!]
\begin{center}
{\includegraphics[width=0.79\textwidth]{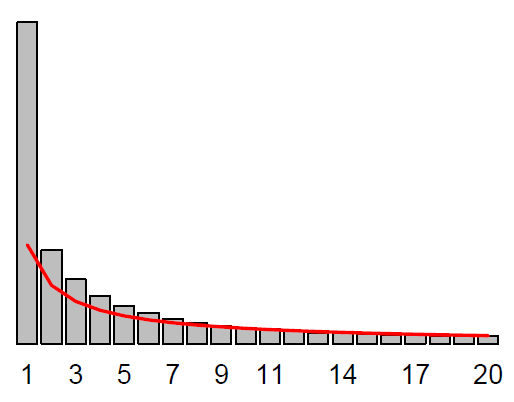}}
\end{center}
\caption{Frequencies of users visiting a particular website during a given period of time and NBD fit for Gamma-Poisson model in presence of the cookies effect (c). }
\label{fig:NBD}
\end{figure}

Mixed Poisson process is a classical probabilistic model used in many different applied areas, where dynamics of occurrence of certain events is of the main interest; see \cite{Grandell}.
In consumer behaviour studies, this model can be interpreted as follows.
There is a large number of users/buyers. Each buyer makes events (visits, conversions, purchases) according to a Poisson process with his/her own intensity $\lambda$.
There is a population (mixing) distribution for $\lambda$. The most popular mixing distribution is Gamma-distribution in which case the model is called Gamma-Poisson and  the number of events during a given time period has the negative binomial distribution (NBD).
Furthermore, the Dirichlet distribution can be used for modelling brand selection within a category. Resulting model is called Dirichlet market.
The basic ideas of this repeat-buying model for consumer  research  have been developed by A.S.C. Ehrenberg in his fundamental book "Repeat-buying: Facts, theory and applications". We have adapted this model for studying the internet consumer behaviour. As events, we use  occasions of a purchase or click or simply  of a  visit to a given website.   The main features  distinguishing  internet consumer markets  with regular ones are: (a) there is much more data available (the total weekly number of visits to gambling or shopping websites could easily exceed millions), (b) there are strong time-of-the-day and time-of-the-week  effects (seasonality)  illustrated on Fig.~\ref{fig:hourly}, and (c) as cookies are not kept long, we have to deal with a very unfavorable effect of not recognizing a returning user/client after some random time passes; see Figures~\ref{fig:cookies} and ~\ref{fig:NBD} for illustration. Moreover,  there is no clear definition of a potential  user (buyer) implying that the concept of the total population is undefined. Note, however, that this issue is rather common and  not specific for internet markets.

Feature (a) implies that the data we analyse is an instance of ``big data'' with many features of big data, such as poor quality, present. We routinely  use  supercomputers for cleansing the data and making various aggregations needed to compute different statistics  needed for building the models,  parameter estimation, testing the model adequacy and forecasting.

Figure~\ref{fig:hourly} illustrated effect (b). In this figure, we show hourly series for the the number of conversions of two different products related to banking service and deliveries in the UK. In both time series, we plot about 6 weeks of data, global peaks are reached on Monday mornings, all global minimima  are reached at nights and we can see low numbers during all weekends (they preceed  high values of conversions). In view of very clear time trends such time series are easy to forecast with singular spectrum analysis (SSA); see \cite{Gol}. For the use of the data affected by feature (a) in mixed Poisson process modelling, we need to remove the time heterogeneity by changing the actual time into a virtual time, where  each  hour contains approximately the same number of events (clicks, visits). In order to do that, we need analytic forms of the forecasts which  SSA is conveniently doing for us. These forecasts are also used for early detection of changes in the dynamics of events: an alarm is made when the actual frequencies significantly differ from the forecasted values.

Figure~\ref{fig:cookies} illustrates effect (c) and shows the average time of survival of cookies for different types of browsers in days. Estimation of these survival rates
are essential for adjusting the mixed Poisson  process model. Fortunately, there is  large quantities of data and these survival rates can be estimated with high accuracy.

Figure~\ref{fig:NBD} illustrates that despite the Gamma-Poisson process model can adequately describe the number of visits to a particular website (in this case,  a shopping website), the frequencies of visits  show larger number of users with small numbers of visits (and hence smaller numbers of loyal users with high numbers of visits), relative to a suitable NBD model. This is caused by our inability of  recognizing certain returning users. Depending on the rate of cookies forgetting, we are able to estimate the the amount of missing loyal customers and consequently  adjust the model.

We use the models based on  mixed Poisson processes for the following purposes:
(i) studying periodic patterns and trends,
(ii) forecasting visits to websites (clicks, sales) for planning marketing activity,
(iii) monitoring stability of (early detection of changes in buying behaviours),
(iv) studying discrepancies in behaviour of different cohorts of users,
(v) modelling loyalty, and
(vi) multivariate time series analysis for studying causality and cross-dependence  between traffics to different websites
and between traffic, clicks and conversions; this includes the use of the  Dirichlet market model for modelling brand competition.

\section*{Impact}
Impact has been achieved through the implementation of the computer algorithms in the activites of a London-based digital marketing company, Crimtan (https://crimtan.com/). Crimtan, with offices in 12 countries, delivers dynamic digital marketing campaigns, specifically in the area of real-time bidding, for a large number of clients all over the world. These include Virgin, American Express, Universal, Sky, Volkswagen, Hilton, Visa and HSBC.

With the data at their fingertips, Crimtan have been able to offer their clients highly-targeted advertisement campaigns. The automation of the real-time bidding process has also allowed Crimtan to streamline its own services, resulting in efficiency gains and a significant annual increase in their turnover.

Improved algorithms and using statistical techniques to improve the efficiency and thought processes behind online advertisements have made significant improvements with the accuracy of statistical techniques and hence reduce costs for Crimtan. In particular, the decision-making statistical tools developed in Cardiff helped to reduce the number of human experts employed to make decisions on how and when particular ads can be shown. Experts work efficiently but they are only humans; on the other hand, fast paced computerised calculations nicely emulate the experts and potentially may have a very significant impact.


\end{document}